\documentclass[twocolumn]{aastex63}

\usepackage{soul,xcolor}
\setstcolor{red}

\submitjournal{ApJ}

\shorttitle{VLBA survey of Orion radio stars}
\shortauthors{Forbrich et al.}

\begin{document}

\title{A VLBA Survey of radio stars in the Orion Nebula Cluster: I. The nonthermal radio population}

\author[0000-0001-8694-4966]{Jan Forbrich}
\affiliation{Centre for Astrophysics Research, University of Hertfordshire,
College Lane, Hatfield AL10 9AB, UK}
\affiliation{Center for Astrophysics $\vert$ Harvard \& Smithsonian, 60 Garden St, Cambridge, MA 02138, USA}

\author[0000-0001-6010-6200]{Sergio A. Dzib}
\affiliation{Max-Planck-Institut f\"ur Radioastronomie, Auf dem H\"ugel 69, D-53121 Bonn, Germany}

\author[0000-0001-7223-754X]{Mark J. Reid}
\affiliation{Center for Astrophysics $\vert$ Harvard \& Smithsonian, 60 Garden St, Cambridge, MA 02138, USA}

\author[0000-0001-6459-0669]{Karl M. Menten}
\affiliation{Max-Planck-Institut f\"ur Radioastronomie, Auf dem H\"ugel 69,
 D-53121 Bonn, Germany}



\begin{abstract}
We present first results of a four-epoch VLBA survey for nonthermal
emission toward all 556 compact radio sources previously identified
in a deep VLA survey of the Orion Nebula Cluster (ONC).  We identify
VLBA counterparts toward an unprecedented 123 sources. Of these, 41 do not have X-ray counterparts, of which 34 also do not display near-infrared counterparts. Since these cannot be explained by extragalactic background sources, this suggests a component of the ONC population of young stellar objects that are too deeply embedded for even X-rays to be detectable. We find pervasive variability and detect even most of the highest-S/N sources in only one out of four epochs. Neither a negative spectral index nor extreme variability in the VLA data is a good predictor of a VLBA detection.
\end{abstract}

\keywords{astrometry --- radiation mechanisms: non-thermal --- stars:formation}


\section{Introduction}

Even very early stages of star formation are associated with high-energy processes (e.g., \citealp{fem99}). In addition to thermal X-ray emission, mostly from coronal hot plasma, young stellar objects (YSOs) have been observed to display both thermal free-free radio emission from ionized material in their vicinity and also non-thermal (gyro)synchrotron emission from electrons gyrating in magnetic fields (e.g., \citealp{gue02}). The latter is akin to scaled-up coronal-type activity, and it is often orders of magnitude more luminous than activity on our Sun. To reach these energies, the magnetic structures involved may be rooted in the YSO, but could also connect the star and its protoplanetary disk.

Observationally, it can be difficult to disentangle free-free and non-thermal emission, since they can arise from small scales, and they will often both occur in the same object. There are four main diagnostics that can be used to identify the emission mechanism. First of all, polarization will indicate gyrosynchrotron (circularly polarized) and synchrotron (linearly polarized) radiation. Then, the radio spectral index $S_\nu \propto \nu^\alpha$ can provide information as well, but it will often be ambiguous. Third, variability can be an indication, with rapid variability a sign of non-thermal emission (on timescales as short as a few minutes, e.g. \citealp{for17}), but there is again some ambiguity on intermediate timescales. Fourth, radio brightness temperature is a rarely employed but powerful indicator, since this technique allows us to select sources with brightness temperatures that are above that of thermal emission. This can naturally be accomplished by using Very Long Baseline Interferometry (VLBI), with which it is thus possible to filter out any thermal emission, enabling a census of nonthermal radio emission that remains largely unaffected by thermal radio emission. Apparent brightness is inversely proportional to the synthesized beam size for an unresolved source, and for milli-arcsecond resolution VLBI observations in C-band (4--8 GHz), even faint detections have corresponding brightness temperatures of several $10^6$~K, well above possible values of free-free emission ($\approx 10^4$ to $10^5$~K), even more so when considering that this is a lower limit. Even at this high resolution, the stellar radio emission remains unresolved, with a synthesized beam size corresponding to a linear scale of order 1  AU in our case. Overall, the two most useful indicators thus are polarization and brightness temperature.

In recent years, it has become possible to study populations of radio stars at unprecedented sensitivity, with a range of radio observatories. Our focus has been on observing YSOs in the Orion Nebula Cluster, the nearest  young rich open cluster containing high-mass stars ($d\sim400$~pc; \citealp{men07,kou17}).  Altogether, the ONC has  3500 stellar members that formed less that 2 Mys ago \citep{hil13}, many of which emit X-rays (e.g., \citealp{get05b}). Radio emission from YSOs in the ONC and the nearby Kleinmann-Low nebula was first observed when the NRAO Very Large Array became available \citep{gar87, chu87}, and prior to the VLA expansion, 77 radio sources had been observed in this cluster \citep{zap04}. After the VLA expansion was completed, we obtained a deep pointing with what is now called the Karl G. Jansky Very Large Array, which allowed us to identify 556 radio sources in the ONC \citep{for16}. Our analysis revealed extreme variability for many objects, strongly suggesting non-thermal emission \citep{for17}. While this analysis included the derivation of in-band spectral indices for the brightest sources, it has been impossible so far to extract reliable wideband polarization information across the entire primary beam of this deep pointing.

To obtain a more reliable census of the incidence of non-thermal emission, and to then use this information for a proper motion census of the inner ONC, we have utilized the upgrade of the NRAO Very Long Baseline Array (VLBA). Crucially, this upgrade not only brought a significantly improved sensitivity, particularly in the C-band, but also the advent of the DiFX software correlator \citep{del07,del11}. Previously, correlation was limited to at most a few sources per experiment (e.g., \citealp{men07}), but with the software correlator, it has become possible to correlate {\it all 556 sources} that we identified in our VLA data, in a near-identical primary beam. It is thus possible to obtain an unbiased dataset for which no assumptions need to be made on which sources are most likely to show non-thermal emission.

In this paper (Paper I) we address the incidence of non-thermal radio emission in 556 VLA-identified sources in the ONC, based on four separate observing epochs. An accompanying second paper (Paper~II; Dzib et al. 2020, {\it subm.}), discusses the proper motion measurements of a subsample of our sources with detections in multiple epochs and how these complement {\it Gaia} results in a region that contains optically invisible deeply embedded objects and parts of which show bright nebulosity. A third paper will discuss variability and polarization information extracted from our dataset.

\section{Observations and data reduction}

We have observed the ONC with the VLBA on four occasions. The first observation was carried out on 2015 October 26 (BF117). In a follow-up program (BF123), three more observations were obtained on 2017 October 26 and 27 and on 2018 October 26. This experiment thus was designed to eliminate the effect of parallax by always observing at the same time of year. Note that we added a repeat epoch in 2017 to evaluate the effects of short-timescale variability (i.e., within a day). Identical observing setups were used, as described below, but antenna availability differed between epochs. For an overview about key parameters of the observations used here, see Table~\ref{tab_obs}.

Observations were carried out in C-band at 2~Gbps in dual polarization with an aggregate bandwidth of 256 MHz in a frequency range of 7.068 to 7.324~GHz, using the Roach board-based Digital Backend with its Polyphase Filter Bank. 16 baseband channels in the upper sideband with 32~MHz bandwidth each were recorded with each band recorded in both LCP and RCP. To improve ionospheric calibration, the observations were interspersed with three sets of dual-frequency (4--8 GHz) geodetic observations (`blocks') of 30~min duration each (see \citealp{reb04} for a related discussion), scheduled at the beginning and end of the observing run and in-between the science observations. The pointing position, 05$^{\rm h}$35$^{\rm m}$14$\fs$479 --05$^{\circ}$22$\arcmin$30$\farcs$57 (J2000), was the same as that used in our deep VLA observations \citep{for16}. The phase calibrator for rapid switching was J0541$-$0541, at an angular distance of 1.6$^\circ$, where 40~sec on the phase reference source were followed by 120~sec on the ONC position. The resulting on-source time was about 4.7~hours in each of the epochs. Due to differences in antenna availability, the $(u,v)$ coverage varies slightly from epoch to epoch, with typical synthesized FWHM beam sizes listed in Table~\ref{tab_obs}. The projected minimum baseline corresponds to about 3~M$\lambda$ ($\sim$125~km), while the maximum baseline length depends on the availability of the antennas at Mauna Kea and/or St Croix.

In terms of spectral coverage, our VLBA observations thus have been recording a subset of the bandwidth that was also recorded in our VLA experiment, albeit at a lower nominal sensitivity. While the concatenated VLA data have a nominal sensitivity of 3~$\mu$Jy\,bm$^{-1}$, often not reached due to remnant emission of the largely filtered out but bright nebula, the nominal sensitivity of the VLBA images is 30~$\mu$Jy\,bm$^{-1}$, but with a synthesized beam size that is $\sim$100 smaller (see below) and no remnant extended structure. This nominal sensitivity varies somewhat from epoch to epoch, not least due to different numbers of available antennas. Given the wider band of the VLA C-band observations, even though it encompasses the frequency range covered here, the primary beam of the corresponding VLA experiment is slightly larger than that of the VLBA experiment, depending on the frequency considered.

While our VLA study has already demonstrated the occurrence of extreme variability in this sample \citep{for17}, we can obtain a few baseline comparisons to estimate how many sources may be detected in the VLBA data. First of all, 508 out of the 557 targets are located within the nominal VLBA primary beam (FWHM) at the observing frequency. Assuming that all of these are both nonthermal (i.e., detectable by VLBI) and non-variable (i.e., at the same flux level as during the VLA observations), which in combination is very unlikely, we would expect 98 sources to be detected above a signal-to-noise ratio, S/N, of 6.5, our sensitivity cut-off (see below). The actual number count will primarily depend on the fraction of VLA sources that have nonthermal emission and show variability. Already the extreme variability exhibited by some of the sources of our sample when observed with the VLA \citep{for17} underscores the limited meaningfulness  of using a time-averaged flux density as an S/N threshold.

We have used the multi-object capabilities of the DiFX software correlator \citep{del07,del11} to obtain VLBA data for all 556 compact VLA sources. Starting with the follow-up experiment BF123, we additionally correlated at the position of COUP~672 as the 557th target, following its detection at radio frequencies by \citet{dzib2017}.

Data reduction was carried out using standard procedures using the NRAO Astronomical Image Processing System\footnote{\url{http://www.aips.nrao.edu/}} (AIPS). In a first step, systematic delays were removed based on measurements of the ionospheric total electron content, as provided for the time of the observation by Global Positioning System data, and all data were corrected using the latest USNO Earth orientation parameters. Subsequently, residual delays from clock drifts and zenith atmospheric delays were measured using the geodetic blocks and removed. This step involved the AIPS routine DELZN to estimate clock delays. Electronic delays and differences among the IF bands were then removed using observations of the strong calibration source J0530+133. Finally, we interpolated the phases of the reference source and applied this identical calibration to all separately correlated science targets. These steps were repeated for all four epochs considered here.

\begin{deluxetable*}{llll}
\tabletypesize{\footnotesize}
\tablenum{1}
\tablecaption{Observing log\label{tab_obs}}
\tablewidth{0pt}
\tablehead{
\colhead{Epoch} & \colhead{Date, Time (UT)} & \colhead{Beam\tablenotemark{1}} & \colhead{Antennas}}
\startdata
BF117  & 2015 Oct 26, 06:25--14:23 & 4.7$\times$ 1.6 & 8 (no MK, HN)  \\
BF123A & 2017 Oct 26, 06:23--14:21 & 4.4$\times$ 1.3 & 9 (no SC, PT\tablenotemark{2}, HN\tablenotemark{2})      \\
BF123B & 2017 Oct 27, 06:19--14:17 & 4.1$\times$ 1.4 & 8 (no SC, HN, PT\tablenotemark{2})  \\
BF123C & 2018 Oct 26, 06:24--14:22 & 2.8$\times$ 1.2 & 10             \\
\enddata
\tablenotetext{1}{Synthesized beam (FWHM), in milliarcseconds}
\tablenotetext{2}{operational, but flagged during processing}

\end{deluxetable*}

\section{Data analysis}

\subsection{Source detection}

To check for source detections, all 557 targets in all four epochs were imaged using AIPS in Stokes~$I$, resulting in more than 2200 images. To search for peaks in a wide field, this first set of images consisted of images of size  2048$^2$ pixels with a pixel size of 0.5~mas, i.e., we mapped areas of about a square arcsecond, centered at the VLA positions, using natural weighting. The smallest synthesized beam of $\sim 1.2\times2.8$~mas was slightly undersampled with at least two pixels per half-power beam width in each axis, and the main goal was to search for detections in a wide area. This resolution corresponds to AU scales at the distance of the ONC, and any coronal-type emission should thus be unresolved. 

For a second round of imaging, we then conservatively selected peaks above a threshold of S/N = 5.5, as determined from the peak pixel value and the pixel rms in the images produced with AIPS, to be imaged in greater detail. We here utilize the work of \citet{her17}, who analyzed the role of noise in the source detection process when analyzing VLBI images of faint radio sources in the COSMOS field. While that study was obtained at a frequency of 1.4~GHz, the COSMOS field lends itself more easily to a systematic study of detection probabilities than the comparatively crowded ONC. These authors found clear signs of non-Gaussian noise when going to lower S/N detection levels (see also \citealp{mid13}). The joint probability of finding a false positive or a chance detection within $0\rlap{.}''4$ of a VLA source was estimated to be 19\%, 0.2\%, and 0.02\% for cut-offs of S/N=5, 5.5, and 6, respectively, clearly a non-Gaussian progression. They concluded that a cut-off at S/N = 7 means that the false-positive detection rate can be ignored. 

In our case, we have a less stringent constraint on the offsets between the VLA and VLBA positions than if we expected constant positions, since significant proper motions may have occurred in our sources during the course of this experiment, and even different components of multiple systems may be detected in some epoch(s), but not in others. Our search area is approximately 2.5 times larger than that of \citet{her17}. We pick a cut-off of S/N=6.5 for an estimated false-positive rate of $\sim$0.1 sources across the full sample and four epochs. The nominal Gaussian expectation at this significance level would be 0.02 false detections, leaving a factor of a few for any non-Gaussian noise, while still ensuring a low false-positive rate.

To illustrate the need for a cut-off at this relatively high S/N level, we note that all images of the first epoch (to give an example) contain peaks with S/N$>$4.6, and 457 maps have peaks with S/N$>$5, clearly pointing at the presence of noise peaks. We also note that the maximum number of independent elements in our images is the number of enclosed FWHM synthesized beams. Our large images of four million pixels typically contain between 120,000 and 270,000 synthesized beam areas. As a result, within our complete sample of 557 targets, we would expect about one spurious detection due to the chance appearance of a noise peak at S/N$>$5.5 by Gaussian noise alone, motivating the choice of a higher cut-off.

For the brightest peaks thus identified we then produced fully sampled images using CASA. A target position identified in one epoch was imaged in all four epochs. Here, we again used image sizes of 2048$^2$ pixels, but with a pixel size of 0.05~mas, i.e., we mapped areas of 0.1$\times$0.1 arcseconds, using natural weighting. A total of 171 source positions showed peaks with S/N$>$5.5 and were thus imaged in this way, with a typical image rms sensitivity of 21~$\mu$Jy\,bm$^{-1}$.

We additionally concatenated the data of the two epochs separated by just one day, BF123A and BF123B, that were only separated by one day to additionally search for detections at higher sensitivity, and a few additional sources beyond those identified in the analysis below were identified. While our discussion is otherwise limited to the individual epochs, we highlight these additional sources in our discussion below.

Several factors complicate the source detection: 1) many sources are likely variable,  particularly given the nonthermal nature of the emission, 2) unknown but significant proper motions between epochs (except for BF123AB) make it impossible to look for coincident detections, 3) the target positions are only known at about 100 times lower resolution with different spatial filtering properties; the VLA detections could split into multiple VLBA sources, and 4) given the overall number of pixels, there is an increased chance of detecting noise peaks as seemingly significant sources. 

Visual inspection of the results quickly highlighted the fact that the detected sources are highly variable, as can be expected from nonthermal emission, which is underlining the value of multiple observing epochs for this experiment. We adapted our detection criteria accordingly by only requiring one detection above the cut-off of S/N$ >$ 6.5.

\section{Results and discussion}

Among 557 targets we report 123 detections, including 12 identified in the concatenated epoch BF123AB. Next to the 386 sources with peaks of S/N$<$5.5 and thus not selected for detailed imaging, the remaining 48 sources out of the imaged sample of 171 sources are considered to be nondetections. Out of 557 targets, we thus have 123 detections and 434 nondetections. The detections are listed in Table~\ref{tbl_AB}, where we also list additional information including on the identification of X-ray and NIR counterparts and on the nominal angular separation of these detections from the VLA positions. The number of detections is greater than expected from scaling the VLA flux densities alone (98), which is probably due to the fact that these flux densities, averaged over almost 30~h, underestimate the flux density observable during flares. 

It is also interesting to consider the positional offset when compared with the nominal VLA positions, based on observations obtained in 2012 \citep{for16}. While a number of positions fall within a few 10 mas of the nominal VLA position (from 2012), more extreme examples include the detection at S/N=7.9 in source 300, at a distance of 580 mas from the expected position, corresponding to about 230~AU at the distance of the ONC, and plausibly a component of a multiple system. This highlights why no cut-off in angular separation has been applied within the imaged areas.

A full discussion of the positions and proper motions of the detected sources is presented in Paper~II. Here, we are primarily interested in quantifying the number of VLBA detections in our large VLA sample. The most striking result, enabled by studying four separate epochs, is the enormous influence of variability, a clear sign of nonthermal emission, on the detection count. Additionally, we detect, on two occasions, {\it two} radio sources toward a single VLA source on VLBA scales (sources 177 and 414), adding to the sample of radio binaries in the ONC, and underscoring the potential for significant positional offsets when compared to the nominal VLA position. These are discussed in Paper~II.

\begin{figure}
\includegraphics[width=\linewidth]{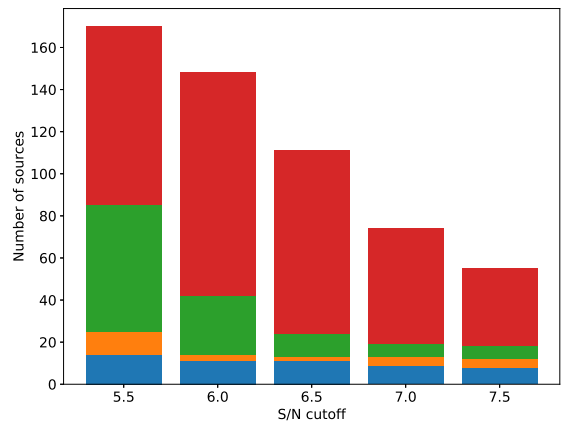}
\caption{Number of sources detected above a given S/N threshold, color-coded by number of detections among four epochs (red=1, green=2, orange=3, blue=4). \label{hist1}}
\end{figure}

We illustrate the impact of variability on this detection experiment by highlighting the number of detections per source among the four epochs, as a function of S/N cut-off. In Figure~\ref{hist1}, we plot the number of nominal detections, defined as peaks in the search area per source, as a function of several different S/N cutoffs, color-coding the {\it number} of detections among the four epochs. Not too surprisingly, the overall number of detections decreases for higher S/N cutoffs. However, even in the highest-S/N group with S/N$>$7.5, most sources reach this level of significance in just one out of four epochs: Out of a total of 55 sources with a maximum S/N per epoch of S/N$>$7.5, 37 reach this level in just one epoch, and only 8 sources reach it in all four epochs. 

A particularly instructive look at the role of variability comes from a comparison of epochs BF123A and BF123B, which were observed on two subsequent days. Out of 28 sources that were detected at S/N$>$6.5 in BF123A, only eleven sources were detected at S/N$>$6.5 on the next day. 

\subsection{Source identification}

As discussed in \citet{for16}, the multi-wavelength identification of the VLA sources is not entirely straightforward. While most of them are expected to be YSOs emitting nonthermal radio emission, the sample also contains sources with thermal radio emission, for example arising from protoplanetary disks externally photoionized by UV radiation from the ONC’s most luminous  O-type star $\theta^1$~Ori C, the so-called {\it proplyds} \citep{ode93}, or from outflows and jets (e.g., \citealp{ang18}).

A correlation with X-ray data is the best way of identifying emission from YSOs, given the reduced extinction cross section of the interstellar medium in the X-ray range and and the lack of source confusion due to nebulosity in the Orion Nebula, which in comparison lowers the sensitivity of infrared observations. Already when discussing the VLA sample, we had correlated the radio sample with the deep X-ray catalog from the Chandra Orion Ultra-deep Project (COUP,  \citealp{get05}) to identify YSOs with a search radius of $0\farcs5$, while keeping in mind that due to variability and extinction, an X-ray non-detection could still be related to a YSO. We use the VISION survey to identify near-infrared counterparts \citep{mei16}. X-ray and infrared non-detections do not necessarily rule out ONC members, if the extinction is very high (affecting both X-rays and infrared emission) or if bright emission from the nebula significantly reduces the sensitivity in the infrared. One example is the flare source from \citet{for08}, from which no infrared emission was detected, and only a brief X-ray flare enabled the identification as a YSO.

One hypothesis would thus be that the VLBI experiment preferentially singles out VLA sources with X-ray counterparts, since this would increase the likelihood of detecting coronal activity. We have included the associated COUP sources from \citet{for16} in Table~\ref{tbl_AB}, keeping in mind that particularly in the case of multiple systems, the X-ray and radio emission may be unrelated. Near-infrared information is included as well.

Interestingly, from among the 123 detections, a considerable subset of 41 sources (or one third) do {\it not} exhibit an X-ray counterpart in the COUP survey. Overall, however, the X-ray incidence rate is two thirds, correspondingly, which is higher than that of the full VLA sample at 46\%. It is worth noting that the highest S/N of an X-ray non-detection occurs in source 11, at S/N = 8.7, and all remaining X-ray non-detections have VLBI detections with a peak S/N ranging from 6.5 to 8.7. This range in detection significance overlaps with that of VLBI sources with X-ray detections, and the non-detections thus are unlikely to be entirely due to the relative sensitivity of {\it Chandra} and the VLBA to detecting YSOs. Near-infrared data do not significantly change this picture, as 34 out of the 41 sources without X-ray counterparts also do not display NIR counterparts while 72 of the 82 sources with X-ray counterparts show near-infrared detections.

Some of the VLBI detections without multi-wavelength counterparts could be extragalactic background sources. However, in the entire VLA C-band beam, we would only expect at most three extragalactic sources at $S>0.195$~mJy (or 6.5 times the typical rms of the observations reported here), based on \citet{win93}. An additional perspective on these questions will be afforded by the analysis of proper motions, but the intriguing conclusion is that most of these targets belong to the YSO population of the ONC even though they so far have remained undetected in X-rays.

While X-ray data provide a means of identifying YSO counterparts to the radio sources, extensive literature exists also on spectral types and YSO evolutionary classes in the ONC, even if the X-ray data arguably provide the most complete census. The main problem is that in various wavelength regimes, the Orion Nebula's bright extended emission precludes the collection of data for various sources, which often means that only limited non-radio band information is available on a given object. In particular, note that only a fraction of the ONC's member stars could be measured by {\it Gaia} (see Paper II). For many Orion sources, there are indications of multiplicity, which impacts this study in two different ways: 1) the attribution of a spectral type to a the radio emitting stellar component may be ambiguous, and 2) positional offsets would be expected if just one component in a multiple system is detected by VLBI.

As discussed in \citet{for16}, the VLA radio sample comprises sources with spectral types ranging from M to O as catalogued most recently by \citet{hil13}, and the same is true for the VLBI sample (when assuming that actual counterparts are compared). Spectral types of 56 of our targets have been reported by \citet{hil13}. Additionally, eight of our VLBI detections are listed, within our search radius of 0.5$''$, in the X-ray/NIR study of the inner ONC by \citet{pri08}, who determined evolutionary classes for the X-ray detected YSOs in the ONC, all T~Tauri stars. Six out of these are listed as class~III sources, and two as class~II sources. However, with fewer than 5\% of even our input VLA sample having counterparts in \citealp{pri08}, an assessment of detection rates in different YSO evolutionary stages cannot be obtained at this time. Spectral types and protostellar classes, where available, are listed in Table~\ref{tbl_AB}.

It is also worth considering constraints from the deep VLA observations that we could have used to select candidate nonthermal targets if it would have been impossible to conduct an unbiased VLBA survey. Interestingly, the VLBA detections reported here do not correlate strongly with observable VLA characteristics, underlining the value of this unbiased survey. Other than previous VLBA detections \citep{men07,kou17}, we could have selected the subset of VLA sources with significantly negative in-band spectral indices in \citet{for16}. As discussed in our VLA paper, out of 170 sources with reliable spectral indices, the best candidates for nonthermal emission are the 17 sources therein that have negative spectral indices with a significance of at least 3$\sigma$. Remarkably, only two of these 17 sources ([FRM2016] 127 and 148) show up in our list of VLBA detections, while overall 53 out of the 170 sources with reliable spectral indices have been detected with the VLBA. Similarly, we could have picked the 13 extremely variable VLA sources (varying by more than an order of magnitude on short timescales) from \citet{for17}, and here the VLBA detection rate is somewhat higher: seven out of these 13 sources ([FRM2016] 53, 98, 189, 254, 319, 414, and 515) are VLBA detections in our experiment, in addition to the flare source from \citet{for08}, which is also detected ([FRM2016] 198). While the detection of nonthermal emission with the VLBA is unambiguous, the intermittent nature of nonthermal emission means that multiple epochs of observations may be required in order to characterize the population of stars that show nonthermal emission, and any observable that is impacted by this variability is not a good predictor for nonthermal emission at a different time. Extreme variability at least appears to be a better indicator of nonthermal emission than negative spectral indices in this case, where the VLA spectral indices are an average out of almost 30 hours of observations, perhaps not sufficiently accounting for the intermittent nonthermal variability.

\section{Summary and conclusions}

We present first results of multi-epoch VLBA follow-up of 556 VLA radio sources in the ONC, involving four epochs obtained over a time interval of three years, while eliminating the role of parallax motions. The first goal is to quantify the nonthermal population as defined by the set of VLBI detections of sources in our VLA catalog. Across four epochs, we find an unprecedented total of 123 detections. We find strong variability, as may be expected from nonthermal coronal-type emission, and very few bright sources are detected in all epochs. The incidence of X-ray counterparts among the VLBI sample, at two thirds, is significantly higher than in the VLA sample, indicating that the VLBI observations preferentially single out young stellar sources. However, 41 of the sources do not have X-ray detections, and 34 of those also do not have near-infrared detections, hinting at a population of radio-detected YSOs that remained undetected in X-ray observations due to variability, extinction, or both. When compared with indications of non-thermal emission derived from our previous deep VLA data, we find that neither in-band spectral indices nor extreme variability would have been a good predictor of our VLBA detections, even if significant variability works better in this regard than the negative spectral indices. This again underlines the value of now being able to conduct unbiased surveys with the VLBA due to the advent of software correlation.

\acknowledgments
We thank the anonymous referee for helpful and constructive comments that lead to an improved presentation of our results. The National Radio Astronomy Observatory is a facility of the National Science Foundation
operated under cooperative agreement by Associated Universities, Inc.

\startlongtable
\begin{deluxetable*}{lrrrrrrl}
\tabletypesize{\footnotesize}
\tablenum{2}
\tablecaption{List of VLBA detections in the Orion Nebula Cluster\label{tbl_AB}}
\tablewidth{0pt}
\tablehead{
\colhead{[FRM2016]} & \colhead{COUP} & \colhead{VISION NIR\tablenotemark{1}} & \colhead{$N_{\rm S/N>6.5}$} & \colhead{$(S/N)_{\rm max}$} & \colhead{Ep.\tablenotemark{2}} & \colhead{Sep. (mas)\tablenotemark{3}} & \colhead{Notes\tablenotemark{4}}}
\startdata
  2     &  107 & 05345597-0523130 & 2 &  61.8 & C & 185 & class III; K1-K4		    \\ 
 10     &  262 & 05350628-0522027 & 1 &  13.9 & C &  26 & class III; K5		    \\
 11     &   -- & --		  & 1 &   8.7 & B & 216 &			    \\
 14 	&  283 & 05350727-0522266 & 1 &   6.8 & A & 523 & M4.5  		    \\
 18     &  338 & 05350968-0523559 & 2 &  15.4 & 7 &  43 &			    \\
 21$^*$ &  343 & 05350977-0523269 & 1 &   8.6 & AB&  12 & class III; K4-M0		    \\
 22     &  342 & 05350977-0521284 & 1 &  18.0 & B &  12 & M0, K5-K7		     \\
 24 	&  350 & 05350990-0523385 & 1 &   6.7 & 7 & 291 &			     \\
 25$^*$	&   -- & --		  & 0 &   6.8 & AB& 299 &			     \\
 30 	&  363 & 05351026-0521571 & 1 &   6.7 & A & 215 & M5.5  		     \\
 32$^*$	&  378 & 05351050-0522455 & 0 &   6.7 & AB&  16 & K6-M2 		     \\
 35$^*$ &  390 & 05351062-0522560 & 0 &   7.1 & AB& 444 &			     \\
 37 	&  394 & 05351073-0523446 & 1 &   7.0 & B &  79 & K2-M0 		     \\
 42     &   -- & 05351093-0523267 & 1 &   7.8 & C & 260 &			     \\
 47$^*$	&   -- & --		  & 0 &   6.7 & AB& 551 &			     \\
 53     &  427 & 05351156-0524481 & 1 &  12.9 & C &  59 &			     \\
 55     &   -- & --		  & 1 &   7.2 & B & 335 &			     \\
 64     &  444 & 05351172-0525128 & 1 &   7.3 & B &  49 &			     \\
 66     &  450 & 05351180-0521493 & 4 &  78.7 & A &   8 &			     \\
 70     &   -- & --		  & 1 &   7.2 & B & 464 &         		     \\ 
 72     &   -- & --		  & 1 &   7.6 & B & 378 &			     \\
 75 	&  465 & 05351212-0524338 & 1 &   7.0 & C & 380 & M1-M2e		     \\
 86$^*$ &   -- & --		  & 0 &   8.1 & AB& 173 &			     \\
 93     &  504 & 05351285-0521340 & 1 &  17.0 & 7 &   5 & class II			     \\
 98 	&  510 & --		  & 1 &   6.9 & 7 & 372 &			     \\
122 	&   -- & --		  & 1 &   6.5 & B & 526 &			     \\
127     &  551 & 05351352-0522196 & 1 &   7.1 & B & 387 & K-M1  		     \\
129     &   -- & --		  & 1 &   7.9 & B & 448 &			     \\
130     &  554 & 05351358-0523552 & 4 & 144.7 & C &  26 &			     \\
133     &   -- & --		  & 1 &   8.2 & B & 199 &			     \\
135 	&   -- & --		  & 1 &   6.9 & A & 405 &			     \\
137     &   -- & --		  & 2 &   7.1 & B & 279 &			     \\
148     &  593 & 05351392-0523202 & 1 &   7.4 & A & 572 &			     \\
149$^*$ &   -- & --		  & 0 &   8.9 & AB& 425 &			     \\
154     &  594 & --		  & 3 &  78.9 & C &  20 &			     \\
158     &  602 & 05351405-0523384 & 2 &   8.2 & A &  19 & M3			     \\
161 	&  598 & --		  & 1 &   6.6 & 7 & 240 &			     \\
167 	&  608 & --		  & 1 &   6.5 & C & 359 &			     \\
170     &   -- & --		  & 1 &   7.1 & C & 520 &			     \\
176 	&   -- & --		  & 1 &   6.6 & C & 546 &			     \\
177     &  625 & --		  & 2 &  26.0 & C &  13 &			     \\
182     &   -- & --		  & 1 &   7.7 & B & 557 &			     \\
184     &  639 & --		  & 4 &  49.9 & C &  16 &			     \\
188 	&   -- & --		  & 1 &   6.6 & A & 476 &			     \\
189     &  640 & --		  & 1 &   8.3 & 7 &  16 &			     \\
196     &  648 & (det)		  & 1 &  15.9 & 7 &  23 & K3-M2 		     \\
197 	&  645 & 05351465-0520424 & 1 &   6.7 & A & 166 & class II			     \\
198     &  647 & --		  & 2 &   9.0 & C &   8 & Orion Radio Burst Source (ORBS) 		     \\
203$^*$ &   -- & 05351472-0522296 & 0 &   8.3 & AB&   9 &			     \\
205 	&   -- & --		  & 1 &   6.7 & 7 & 522 &			     \\
211     &  662 & --		  & 4 &  24.5 & A &  16 &			     \\
212     &  670 & 05351492-0522392 & 1 &  35.1 & 7 &  14 & K3-M2 		     \\
222     &  689 & 05351526-0522568 & 1 &   7.4 & B & 453 & G6-K7 		     \\
227 	&   -- & --		  & 1 &   6.8 & B & 176 &			     \\
230 	&   -- & --		  & 1 &   6.8 & C & 472 &			     \\
232     &   -- & --		  & 1 &   7.5 & C & 309 &			     \\
240     &  717 & 05351552-0523374 & 1 &   7.0 & 7 & 122 & K			     \\
241     &  718 & (det)		  & 3 &  59.1 & C &  54 & K4-M1 		     \\
242$^*$	&   -- & --		  & 0 &   6.8 & AB& 233 &			     \\ 	   
249   	&  734 & 05351576-0523384 & 1 &   6.6 & C & 514 &			     \\
250     &  732 & (det)		  & 4 &  80.6 & C &  10 & B5-B8+G0-G5		     \\
254     &  745 & 05351582-0523143 & 4 & 214.6 & C &  37 & O9-B1.5		     \\
273     &   -- & 05351608-0523278 & 1 &   7.3 & C & 299 &			     \\
285$^*$ &  783 & 05351619-0521323 & 0 &   7.7 & AB& 508 & M4.5  		     \\
300     &  801 & 05351638-0524032 & 1 &   7.9 & 7 & 580 & K4-K7 		     \\
303     &  806 & 05351642-0522121 & 1 &   7.0 & B & 271 &			     \\
314     &   -- & --		  & 2 &   7.7 & 7 & 496 &			     \\
319     &  828 & 05351676-0524042 & 4 &  11.9 & 7 &  23 & K2-K6 		     \\
321     &  827 & 05351677-0523280 & 1 &   7.6 & A & 539 &			     \\
326     &  841 & 05351693-0522098 & 1 &  10.8 & B & 429 &			     \\
327     &   -- & --		  & 1 &   7.3 & 7 & 447 &			     \\
335 	&  856 & 05351706-0523397 & 1 &   6.8 & A & 345 & K5e			     \\
339 	&   -- & 05351712-0524344 & 1 &   6.8 & C & 139 &			     \\
343     &  867 & 05351721-0521317 & 1 &   8.2 & 7 &  21 & K3-K7 		     \\
347 	&  876 & 05351735-0522357 & 1 &   6.9 & C & 360 & K0-K2,K2-K4		     \\
350     &  874 & 05351739-0522036 & 2 &  14.6 & C &  16 &			     \\
357$^*$ &  896 & 05351750-0521062 & 0 &   7.7 & AB& 162 &			     \\
360 	&  897 & 05351755-0521455 & 1 &   6.7 & B & 420 &       		     \\ 
364 	&   -- & 05351768-0523410 & 1 &   7.0 & B & 277 & K-M			     \\
367 	&   -- & 05351773-0524437 & 1 &   6.9 & A & 611 &       		     \\ 
373$^*$	&   -- & --		  & 0 &   6.8 & AB& 277 &			     \\
378     &  932 & 05351794-0522455 & 4 & 153.9 & C &  13 & G-K3  		     \\
382     &  942 & 05351803-0522054 & 1 &  15.6 & C &  37 & G-M2  		     \\
389     &  956 & 05351820-0523359 & 1 &   7.6 & 7 & 115 & G4-K3 		     \\
398 	&   -- & --		  & 1 &   6.5 & C & 514 &			     \\
400     &  965 & 05351836-0522374 & 4 & 204.6 & C &  17 & G8-M2 		     \\
402 	&  963 & 05351839-0520204 & 1 &   6.6 & C &  36 & class III; K8-M1.5,M1.5-M3,M3-M4 \\
408 	&   -- & --		  & 1 &   6.7 & B & 663 &			     \\
414     &  985 & 05351866-0520337 & 4 &  15.5 & C &  27 & F8-K4 		     \\
426 	&   -- & --		  & 1 &   6.7 & B & 645 &			     \\
435     & 1023 & 05351921-0522507 & 1 &  12.4 & B &  12 &           		     \\ 
440     &   -- & --		  & 1 &   7.1 & C & 281 &			     \\
456     &   -- & --		  & 1 &   7.4 & B & 503 &			     \\
457 	& 1071 & 05352005-0521059 & 1 &   6.7 & A & 332 & K			     \\
459     & 1083 & 05352013-0521336 & 1 &  21.3 & 7 &  20 & class III;$<$M0		     \\
462     & 1087 & 05352016-0526390 & 1 &  16.7 & C & 130 & mid-G-early-K 	     \\
466     & 1090 & 05352021-0520569 & 2 &  39.3 & B &  57 & F7-K3 		     \\
467     & 1091 & 05352027-0525040 & 1 &   7.2 & B & 560 &			     \\
468     & 1100 & 05352039-0522136 & 1 &  11.7 & C &  21 & class III; M0,M1		     \\
470 	&   -- & --		  & 1 &   6.5 & C &  86 &			     \\
477 	& 1110 & 05352064-0522455 & 1 &   6.8 & C & 372 & M4.5  		     \\
480     & 1116 & 05352072-0521443 & 2 &  22.1 & C &   6 & B			     \\
485     & 1130 & 05352104-0523490 & 4 &  36.2 & B &  38 & G8-K5 		     \\
501 	& 1184 & 05352209-0524328 & 1 &   6.7 & A & 579 &			     \\
508 	& 1205 & 05352232-0524142 & 1 &   6.9 & C & 413 & K5-M1.5		     \\
509 	&   -- & --		  & 1 &   6.7 & C & 367 &			     \\
512     &   -- & --		  & 1 &   7.7 & B & 527 &			     \\
514     &   -- & 05352283-0525476 & 1 &   7.7 & B & 234 & M5.5  		     \\
515 	& 1232 & 05352289-0524578 & 1 &   6.6 & C &  80 &       		     \\ 
520     & 1249 & 05352349-0520017 & 1 &   9.3 & C &  64 &			     \\
521     & 1262 & 05352359-0525264 & 1 &   7.3 & 7 & 372 & M3-M3.5e		     \\
522     & 1260 & 05352365-0523463 & 1 &   8.9 & B & 314 & M1			     \\
525     & 1275 & 05352398-0525098 & 1 &   7.8 & A & 561 & M1			     \\
526     & 1276 & 05352403-0523138 & 1 &   7.2 & 7 & 155 &			     \\
527     & 1281 & 05352426-0525187 & 1 &   7.6 & B & 315 & K0-K6 		     \\
530 	& 1289 & 05352447-0524010 & 1 &   6.9 & C & 148 & M1-M3:		     \\
534     &   -- & --		  & 1 &   7.8 & B & 440 &			     \\
535 	& 1313 & 05352503-0524384 & 1 &   6.7 & C & 315 &			     \\
537     & 1311 & 05352508-0523467 & 1 &   9.8 & A &  79 & K2-K5 		     \\
547     & 1360 & 05352639-0525007 & 1 &  14.8 & B &  98 & B0-B2 		     \\
552     & 1428 & --		  & 1 &   7.5 & B & 208 &			     \\
555 	& 1473 & 05353143-0525162 & 1 &   6.6 & A & 422 & B3-B6 		     \\
(557)   &  672 & 05351494-0523393 & 2 &  10.6 & B &  -- & K5-M2 		     \\
\enddata
\tablenotetext{1}{From \citet{mei16}, unlisted detections (e.g., due to crowding) listed as '(det)'.}
\tablenotetext{2}{Epoch of $(S/N)_{\rm max}$: BF117 (7), BF123ABC (A,B,C), or the concatenated epoch BF123AB (AB).}
\tablenotetext{3}{Separation between detection and nominal VLA position, in milliarcseconds}
\tablenotetext{4}{Previous identifications: YSO classes from \citet{pri08} and spectral types summarized from \citet{hil13}. Note that these are for counterparts within 0.5$''$ and do not necessarily represent counterparts of the radio sources, particularly in the case of multiple systems. See Paper II for detailed discussion.}
\tablecomments{Source positions are presented in Paper II. Sources with an asterisk only fulfill the selection criterion in the concatenated epoch BF123AB. The listing of peaks above S/N threshold does not include this concatenated epoch. All sources have four observations each, except for source 557 (see text), which was not correlated in the first epoch (BF117) and which is not listed in \citet{for16}. Source 198 is the ``Orion Burst Source" in the Kleinmann Low nebula, for which a strong outburst at 22.2 GHz was reported by \citep{for08}.}
\end{deluxetable*}


\bibliography{orion}{}
\bibliographystyle{aasjournal}




\end{document}